\documentclass[journal=nalefd,manuscript=letter]{achemso}

\usepackage[version=3]{mhchem} 

\usepackage[psamsfonts]{amssymb}
\usepackage{amsmath}
\usepackage{natbib}
\usepackage{graphicx}
\usepackage{mathrsfs}

\author{Christos~Tserkezis}
\affiliation[Department of Photonics Engineering, Technical University of Denmark]
{Technical University of Denmark}
\email{ctse@fotonik.dtu.dk}
\author{Johan~R.~Maack}
\affiliation[Department of Photonics Engineering, Technical University of Denmark]
{Technical University of Denmark}
\author{Zhaowei~Liu}
\affiliation[Department of Electrical and Computer Engineering, University of California, San Diego]
{University of California, San Diego}
\author{Martijn~Wubs}
\affiliation[Department of Photonics Engineering, Technical University of Denmark]
{Technical University of Denmark}
\alsoaffiliation[Center for Nanostructured Graphene, Technical University of Denmark]
{Center for Nanostructured Graphene, Technical University of Denmark}
\author{N.~Asger~Mortensen}
\affiliation[Department of Photonics Engineering, Technical University of Denmark]
{Technical University of Denmark}
\alsoaffiliation[Center for Nanostructured Graphene, Technical University of Denmark]
{Center for Nanostructured Graphene, Technical University of Denmark}
\email{asger@mailaps.org}
\title[Robust plasmonic nonlocal effects in ensemble measurements]
  {Robustness of the far-field response of nonlocal plasmonic ensembles}

\abbreviations{IR,NMR,UV}

\begin{document}

\begin{tocentry}





\includegraphics[width=0.77\linewidth]{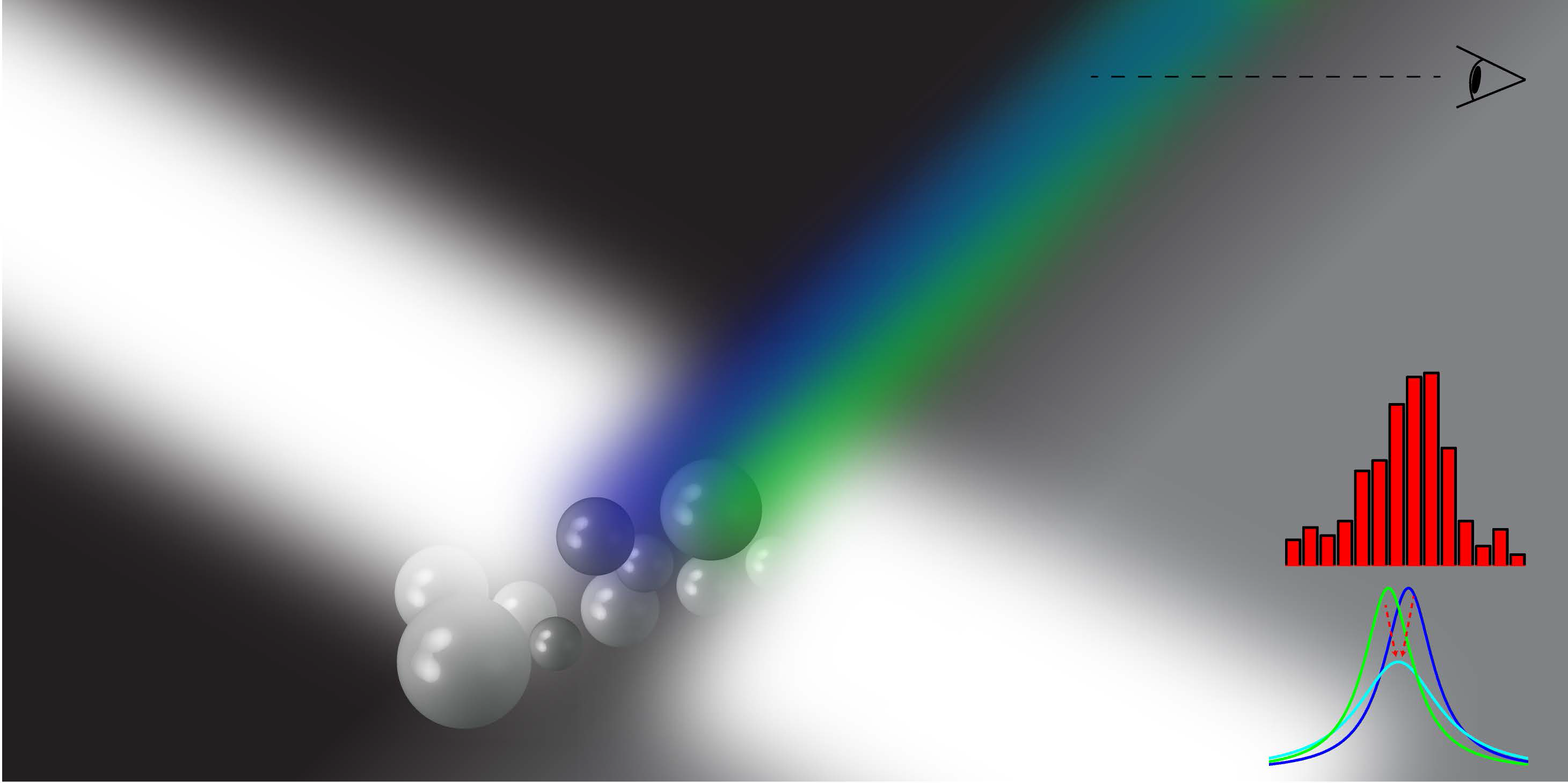}

\end{tocentry}

\begin{abstract}
Contrary to classical predictions, the optical response of few-nm plasmonic particles depends on particle size due to effects such as nonlocality and electron spill-out. Ensembles of such nanoparticles (NPs) are therefore expected to exhibit a nonclassical inhomogeneous spectral broadening due to size distribution. For a normal distribution of free-electron NPs, and within the simple nonlocal Hydrodynamic Drude Model (HDM), both the nonlocal blueshift and the plasmon linewidth are shown to be considerably affected by ensemble averaging. Size-variance effects tend however to conceal nonlocality to a lesser extent when the homogeneous size-dependent broadening of individual NPs is taken into account, either through a local size-dependent damping (SDD) model or through the Generalized Nonlocal Optical Response (GNOR) theory. The role of ensemble averaging is further explored in realistic distributions of noble-metal NPs, as encountered in experiments, while an analytical expression to evaluate the importance of inhomogeneous broadening through measurable quantities is developed. Our findings are independent of the specific nonclassical theory used, thus providing important insight into a large range of experiments on nanoscale and quantum plasmonics.
\end{abstract}


Plasmonics lies among the most prominent research fields in modern nanotechnology,\cite{Maier:2007,Brongersma:2015,Baev:2015} promising exciting applications and unraveling new phenomena as the length scale reduces.\cite{Editorial:2015,Brongersma:2015b,Koenderink:2015} Traditionally, noble metals  constitute the material basis for novel plasmonic devices operating in the visible region,\cite{Murray:2007} although many efforts are recently devoted to extensions towards the ultraviolet, infrared and THz parts of the spectrum.\cite{Naik:2013} A key issue in noble-metal plasmonics is its association with pronounced homogeneous broadening due to Ohmic losses in the metal\cite{Khurgin:2015} and enhanced Landau damping near the surface.\cite{Yan:2015,jin:2015} Within classical electrodynamics, and in the quasistatic regime, radiation losses are small and the limited quality factor of plasmon resonances reflects material losses.\cite{Wang:2006} In other words, homogeneous broadening is important. Furthermore, the commonly employed local-response approximation (LRA) of classical electrodynamics predicts size-independent resonances for the nowadays experimentally accessible small NPs in the quasistatic regime.\cite{Bohren:1983} As a consequence, despite the increasing impact of plasmonics and the promotion of single-particle spectroscopy,\cite{Olson:2015} little, if any, emphasis has been placed on the role of inhomogeneous broadening due to size distribution --- even in experiments on NP ensembles with a noticeable size variation.

The observation of size-dependent resonance shifts not anticipated from classical electrodynamics has recently renewed interest in plasmons in the sub-10-nm regime.\cite{Scholl:2012,Raza:2013a,Raza:2015c} State-of-the-art experiments range from single-particle spectroscopy with the aid of tightly focused electron beams,\cite{Scholl:2012,Wiener:2013,Raza:2013a,Raza:2015c} to optical far-field measurements sampling the response of NP ensembles.\cite{Baida:2009,Taylor:2011,Grammatikopoulos:2013,Ferry:2014,Tserkezis:2014} In the latter case, nonlocal effects\cite{Mortensen:2014,Raza:2015c} and the concomitant inhomogeneous broadening can prove important for the interpretation of ensemble measurements. Ensemble averaging effects have been theoretically explored for exciton systems,\cite{Ruppin:1989} and for large-NP plasmonic collections dominated by retardation-driven redshifts,\cite{Pelton:2013} but related studies in nonlocal plasmonics are still missing. The unambiguous observation of size-dependent resonance shifts in single-particle spectroscopy\cite{Ouyang:1992,Scholl:2012,Raza:2013a,Raza:2015c} encourages therefore to explore broadening phenomena related to size distribution: \emph{What is the robustness of plasmonic nonlocal effects when subject to ensemble averaging?}

The effect of ensemble spectral averaging on the far-field response of nonlocal plasmonic NP collections is studied here theoretically, starting with the ideal case of a normal distribution of free-electron, Drude-like nanospheres. Complexity is subsequently increased by considering more realistic distributions, resembling experimental histograms, of noble-metal NPs, for which additional loss mechanisms like interband transitions and electron quantum confinement are important (the latter affects Drude NPs as well). Through detailed simulations within the framework of Mie theory and its appropriate extensions,\cite{Bohren:1983,Ruppin:1975,Kreibig:1985,Mortensen:2014} we show that ensemble averaging can have significant implications in more ideal cases, but becomes practically negligible in noble-metal plasmonics, which is dominated by homogeneous broadening. Our findings are therefore expected to provide additional flexibility to the design and analysis of experiments on the nanoscale: On the one hand, analyzing the far-field response of a NP collection on the basis of the ensemble mean size is proven sufficient for the purposes of most experimental studies. On the other hand, nonlocal effects are not concealed by single-NP losses in large ensembles, thus allowing to connect with single-particle electron-energy-loss studies.\cite{Scholl:2012,Raza:2015c} 

We first revisit the optical response of a small metallic nanosphere, embedded in air for simplicity. Out study is based on Mie theory\cite{Bohren:1983} and its appropriate extension for nonlocal effects (see Supporting Information).\cite{Ruppin:1973,David:2011,Christensen:2014} The metal is described as a free-electron plasma with transverse ($\varepsilon_{\mathrm{t}}$) and longitudinal ($\varepsilon_{\mathrm{l}}$) dielectric function components given by the frequency- ($\omega$) and wavevector- ($\mathbf{q}$) dependent Drude\cite{Bohren:1983} and hydrodynamic\cite{McMahon:2009,Raza:2015a} models, respectively
\begin{equation}\label{Eq:Drude}
\varepsilon_{\mathrm{t}} (\omega) = \varepsilon_{\infty} - \frac{\omega_{\mathrm{p}}^{2}}{\omega \left(\omega + \mathrm{i} \gamma \right)}~, \quad
\varepsilon_{\mathrm{l}} (\omega, \mathbf{q}) = \varepsilon_{\infty} - \frac{\omega_{\mathrm{p}}^{2}}{\omega \left(\omega + \mathrm{i} \gamma \right) - \beta^{2} q^{2}}~,
\end{equation}
where $\omega_{\mathrm{p}}$ is the plasma frequency of the metal, $\varepsilon_{\infty}$ is the background contribution of bound electrons and ions, $\gamma$ is the damping rate, and $\beta$ the hydrodynamic parameter.\cite{Raza:2015a} We take $\varepsilon_{\infty} = 1$ and $\gamma = 0.01 \omega_{\mathrm{p}}$ to focus on the role of free electrons and ensure low loss (associated with homogeneous broadening). We further assume $\beta = \sqrt{3/5} v_{\mathrm{F}}$ as obtained within the Thomas--Fermi theory,\cite{Raza:2015a} where $v_{\mathrm{F}}$ (= $1.39 \times 10^{6}$ m s$^{-1}$ here) is the Fermi velocity of the metal.

\begin{figure}[ht]
\includegraphics[width=1.0\columnwidth]{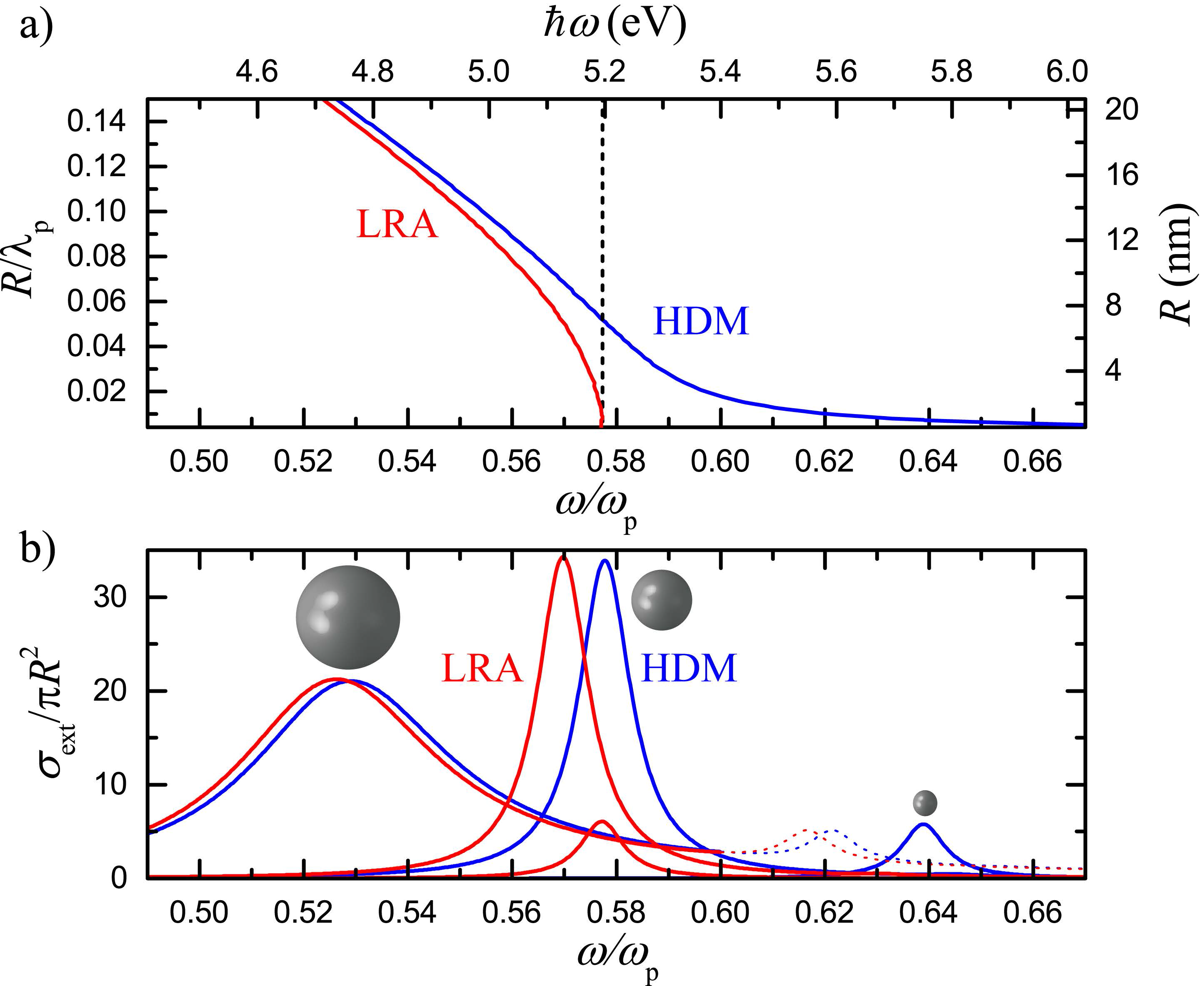}
\caption{(a) Normalized frequency ($\omega / \omega_{\mathrm{p}}$) position of the dipolar plasmonic peak of a spherical NP described by the Drude model of eq~\ref{Eq:Drude} in air, as a function of its normalized radius $R/\lambda_{\mathrm{p}}$, obtained within the LRA (red line) and HDM (blue line) models. The black dashed line displays the prediction of the quasistatic approximation, $\omega_{\mathrm{p}} /\sqrt{3}$. The corresponding energy in eV and radius in nm are given at the top and right axis respectively, assuming a plasmon energy $\hbar \omega_{\mathrm{p}} = 9$ eV. 
(b) Extinction cross section ($\sigma_{\mathrm{ext}}$) spectra (normalized to the geometrical cross section $\pi R^{2}$) for the NP of (a), for three radii, $R/\lambda_{\mathrm{p}} = 0.145$, $R/\lambda_{\mathrm{p}} = 0.051$, and $R/\lambda_{\mathrm{p}} = 0.007$ (from left to right) within the LRA (red lines) and HDM (blue lines) models. For $\hbar \omega_{\mathrm{p}} = 9$ eV these radii correspond to 20, 7, and 1 nm, respectively. The quadrupolar mode of the largest NP is depicted by thin dotted lines.}
\label{Fig:1}
\end{figure}

The size dependence of the frequency of the first (dipolar) plasmonic mode sustained by such a metallic nanosphere of radius $R$ is plotted in Figure~\ref{Fig:1}a as obtained within the LRA ($\omega_{\mathrm{LRA}}$, red line) and HDM ($\omega_{\mathrm{HDM}}$, blue line) models. To make our results scalable for different materials, $\omega$ and $R$ are normalized to the plasma frequency and wavelength, $\omega_{\mathrm{p}}$ and $\lambda_{\mathrm{p}} = 2\pi c/ \omega_{\mathrm{p}}$ respectively. For a better illustration of the sizes and energies usually encountered, the corresponding plasmon energy (NP radius) is provided at the top (right) axis, assuming a typical value $\hbar \omega_{\mathrm{p}} = 9$ eV.\cite{Raza:2015a} For very small NP sizes, LRA reproduces the quasistatic result, $\omega_{\mathrm{LRA}} = \omega_{\mathrm{p}}/\sqrt{3}$ (vertical dashed line in Figure~\ref{Fig:1}a). For larger sizes, retardation causes the modes to drastically redshift and become wider, as also observed in the normalized extinction ($\sigma_{\mathrm{ext}}$) spectra  of Figure~\ref{Fig:1}b (red lines corresponding to different NP sizes within LRA). Higher-order modes will not concern us here, and the quadrupolar plasmon peak of the largest sphere in Figure~\ref{Fig:1}b is only shown by thin dotted lines. The small-size modal frequency saturation predicted by LRA gives place to a continuous blueshift when the metal nonlocal response is taken into account. Comparison between LRA and HDM (blue lines in Figure~\ref{Fig:1}b) immediately shows that the frequency shifts become larger as the NP size decreases, but no additional resonance broadening due to nonlocality is observed. 

A significantly different behavior is expected in a statistical ensemble of small particles, where the strongly blueshifting modes of single NPs will overlap in a sequential manner, leading to important line broadening possibly even for  narrow size distributions, in analogy to the effect of retardation on large NPs.\cite{Pelton:2013} At this point we should also note  that for the type of Drude metal described here, more detailed theories based on atomistic \emph{ab initio} calculations predict frequency redshifts of similar magnitude, instead of blueshifts, due to electron spill-out.\cite{Stella:2013a,Teperik:2013a,Toscano:2015a,Yan:2015} Indeed redshifts are measured for simple metals such as sodium.\cite{Toscano:2015a} Yet in real noble metals such as silver and gold, the spill-out is less extended and the measured size-dependent blueshifts are well reproduced by HDM. An exact description of a specific material is beyond the scope of this paper, and simple nonlocal models should suffice for the study of ensemble averaging, regardless of the direction and origin of modal shifts.

\begin{figure}[ht]
\includegraphics[width=1.0\columnwidth]{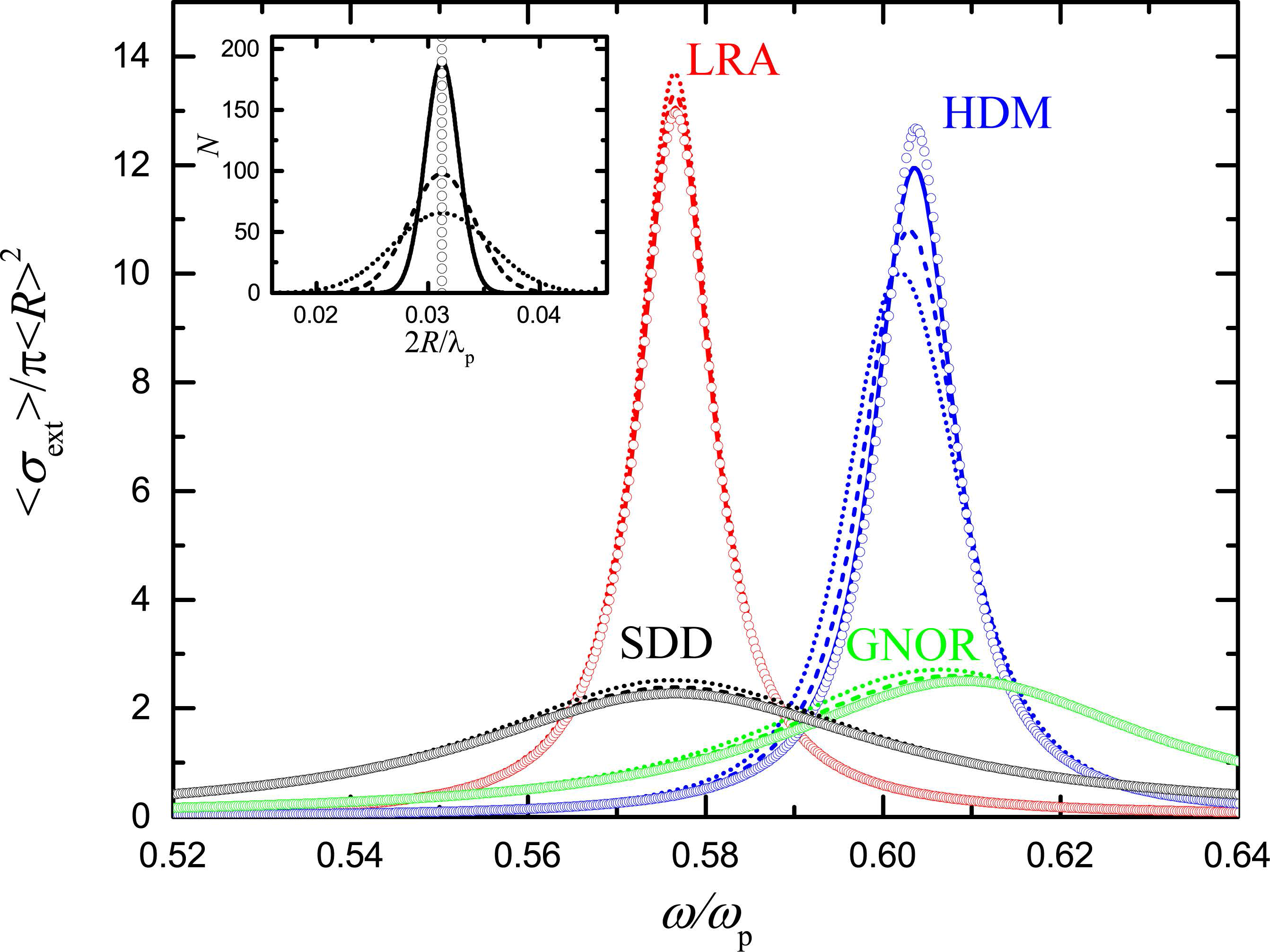}
\caption{Averaged normalized extinction ($\langle \sigma_{\mathrm{ext}} \rangle$) spectra calculated for $N =$ 1000 NPs described by the dielectric function of eq~\ref{Eq:Drude} within the LRA (red lines), HDM (blue lines), GNOR (green lines) and SDD (black lines) models, for the size distributions shown in the inset. The average NP diameter is $2 \langle R \rangle/\lambda_{\mathrm{p}} = 0.031$, which for $\hbar \omega_{\mathrm{p}} = 9$\,eV corresponds to 4.3 nm, and the standard deviation of the normal distribution function is 0.2 (solid lines), 0.4 (dashed lines), and 0.6 (dotted lines). Open circles denote the corresponding spectra for the single $\langle R \rangle$-NP, corresponding to the $\delta$ function distribution (open circles) of the inset.}
\label{Fig:2}
\end{figure}

Ensemble spectral averaging is at a first step investigated by considering a collection of $N = $1000 of the NPs described above, with a mean diameter $2\langle R \rangle /\lambda_{\mathrm{p}} = 0.031$ (corresponding to 4.3 nm for $\hbar \omega_{\mathrm{p}} = 9$ eV). The NP size follows normal distributions around this mean value as shown in the inset of Figure~\ref{Fig:2}, with standard deviations ranging from 0.2 (narrowest distribution, solid line) to 0.4 (dashed line) and 0.6 (widest distribution, dotted line). The extreme case of a $\delta$-function distribution, i.e., all NP diameters corresponding precisely to the mean value, is depicted by open dots. This kind of $\delta$-function distribution is exactly what one assumes in practice when disregarding ensemble averaging. We also note that, while the distributions of Figure~\ref{Fig:2} are continuous functions, discrete size steps are taken in the simulations, small enough to achieve convergence of the averaged spectra. Apart from the LRA and HDM models, we also discuss calculations based on the commonly employed local SDD model\cite{Kreibig:1969} and the GNOR theory.\cite{Mortensen:2014} Within SDD, the damping parameter $\gamma$ becomes size dependent, $\gamma \rightarrow \gamma + A v_{\mathrm{F}}/R$, to effectively take into account the experimentally observed single-NP damping.\cite{Kreibig:1985} The constant $A$, usually taken equal to 1 (as we do here) although a large range of values can be found in literature, is introduced to phenomenologically describe the reduction of the free-electron path length and to account to some extent for quantum-size corrections in very small NPs.\cite{Kawabata:1966,Kreibig:1969,Kraus:1983,DelFatti:2000,Baida:2009} On the other hand, GNOR reproduces  size-dependent damping in a more physical way, by taking electron diffusion into account as a measure of a variety of electron-scattering effects, including Landau damping due to generation of electron-hole pairs.\cite{Li:2013} In practice one replaces $\beta^{2}$ in eq~\ref{Eq:Drude} with $\beta^{2} + D \left(\gamma - \mathrm{i} \omega \right)$, where $D$ is the diffusion constant of the metal, $D \simeq v_{\mathrm{F}}^{2}/\gamma$.\cite{Mortensen:2014,Raza:2015a} The strength of GNOR is that, for arbitrarily shaped plasmonic NPs, it reproduces both the size-dependent blueshifts and the damping of plasmon modes by a simple correction in the dynamics of the free-electron fluid of HDM, whereas SDD models only capture the damping effects.

With these models at hand, we study in Figure~\ref{Fig:2} how spectral averaging compares to single-NP response. Clearly, for the local models (LRA and SDD, red and black lines respectively), averaging does not practically affect the spectra. For all size distributions, the average extinction $\langle \sigma_{\mathrm{ext}} \rangle$, normalized to the geometrical cross section of the mean-size NP ($\langle R \rangle$-NP), $\pi \langle R \rangle^{2}$ (which is known in experiments), reproduces almost perfectly the spectrum of the single $\langle R \rangle$-NP, without frequency shifts or line broadening. Comparison with Figure~\ref{Fig:1} shows that, in the size range of interest, local theories have already reached the quasistatic limit and the plasmon frequencies do not shift further, thus explaining the behavior of the calculated spectra. The case becomes much different however when the spectra are size-dependent because of nonlocality, as is particularly pronounced by the HDM results. The incomplete spectral overlap for NPs of different sizes leads to a clear broadening of the plasmon peaks, larger as the size distribution becomes wider. In addition, since larger NPs are characterized by larger extinction values, the overlap between large and small particles leads to a decrease of $\langle \sigma_{\mathrm{ext}} \rangle$, and to a gradual redshift of the ensemble resonance comparing with the single nonlocal $\langle R \rangle$- NP. One may therefore conclude that statistical averaging can lead to significant deviations in experimental far-field measurements on ensembles of plasmonic NPs with wide size distributions. Nevertheless, since HDM disregards size-dependent damping mechanisms, it is crucial to take such effects into account. In view of the previous discussion, this is straightforward within GNOR (green spectra in Figure~\ref{Fig:2}). The differences between single-NP and ensemble response are now smoothened, leading to smaller additional modal shifts and almost negligible line broadening due to size inhomogeneity: the spectral width is mainly due to single-particle nonlocal broadening.

\begin{figure}[ht]
\includegraphics[width=1.0\columnwidth]{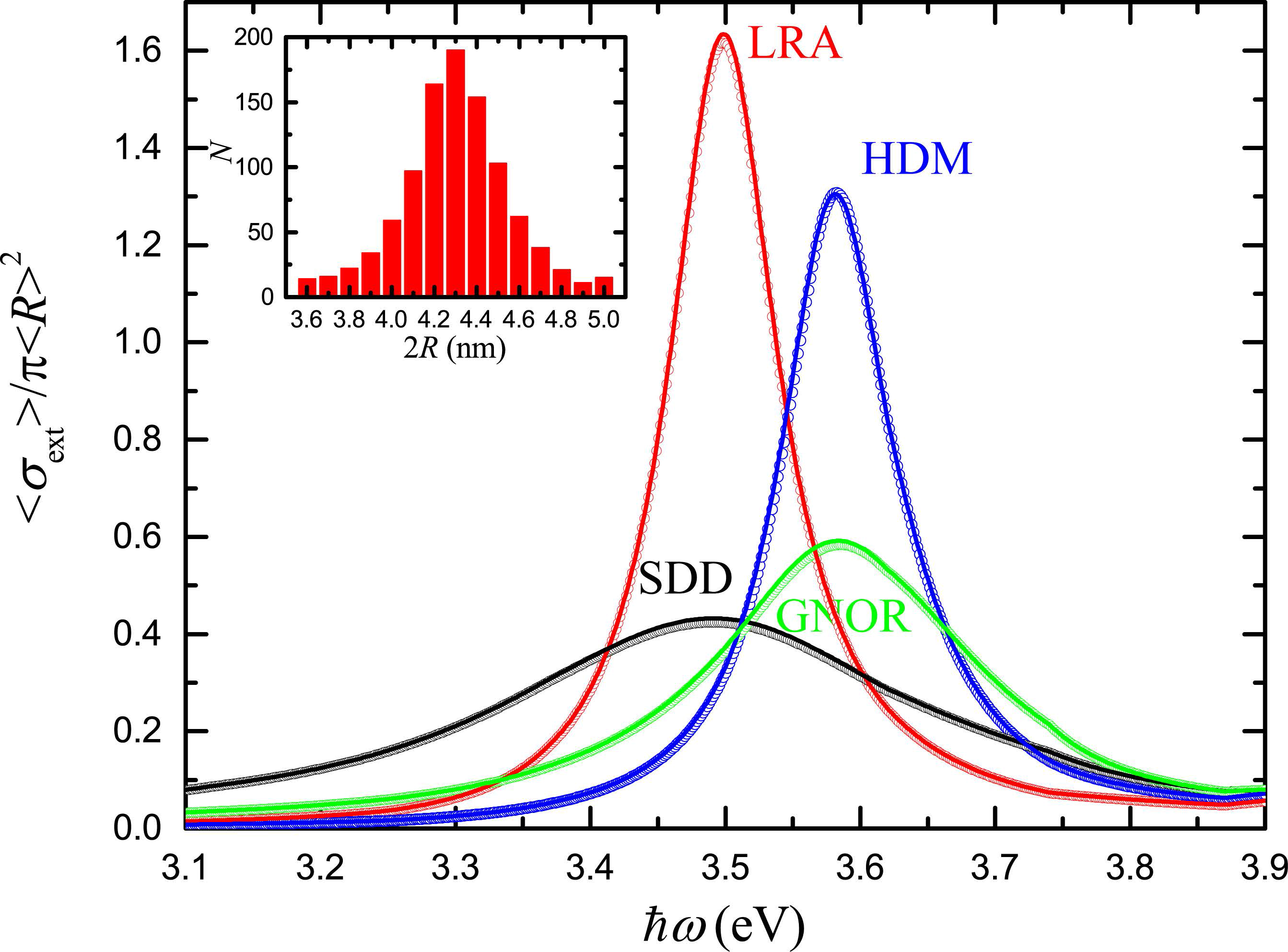}
\caption{Averaged normalized extinction ($\langle \sigma_{\mathrm{ext}} \rangle$) spectra calculated for $N = $1000 silver NPs described by the experimental dielectric function of Johnson and Christy\cite{Johnson:1972} within the LRA (red line), HDM (blue line), GNOR (green line), and SDD (black line) models, for the size distribution shown by the histogram of the inset. The mean NP diameter is $2 \langle R \rangle = 4.3$ nm. Open circles denote the corresponding spectra for a the single $\langle R\rangle$-NP.}
\label{Fig:3}
\end{figure}

The important result of negligible effect of spectral averaging when single-NP size-dependent damping is taken into account may be appealing, but its validity was displayed only for ideal Drude metals and for normal size distributions. In order to connect with more practical, experimentally feasible situations, it is therefore important to carry out similar statistical studies for more realistic distributions in noble metals. We consider a collection of $N = $1000 silver NPs, described by the experimental dielectric function ($\varepsilon_{\mathrm{exp}}$) of Johnson and Christy,\cite{Johnson:1972} following the size distribution shown by the histogram of the inset of Figure~\ref{Fig:3}. In order to apply the HDM, SDD and GNOR models, we obtain $\varepsilon_{\infty}$ in eq~\ref{Eq:Drude} from the experimental values by subtracting the Drude part: $\varepsilon_{\infty} = \varepsilon_{\mathrm{exp}} + \omega_{\mathrm{p}}^{2}/[\omega(\omega + \mathrm{i} \gamma)]$, taking $\hbar \omega_{\mathrm{p}} = 8.99$ eV and $\hbar \gamma = 0.025$ eV, values which describe bulk silver excellently. For SDD and GNOR we further assume $A = 1$ and $D = 3\sqrt{10} A v_{\mathrm{F}}^{2}/(5 \omega_{\mathrm{p}})$, respectively.\cite{Raza:2015a} The calculated spectra of Figure~\ref{Fig:3} display now an almost negligible difference between single-NP and averaged spectra, even for the more pronounced in Figure~\ref{Fig:2} HDM case. Homogeneous line broadening dominates the ensemble optical response, especially when single-NP size-dependent damping is taken into account within GNOR. This observation further strengthens our conclusion that inhomogeneous line broadening is not pronounced in most realistic NP ensembles (despite the non-negligible nonlocal response). Far-field optical experiments on small-NP ensembles can indeed be conducted for the observation of nonlocal frequency shifts, and their interpretation can be performed on the basis of the properties of the $\langle R \rangle$-NP in the collection.

Having considered situations where inhomogeneous broadening can be either strong or negligible, a simple way to decide on its importance without resorting to detailed simulations is desirable. To this end, we develop an analytical model which describes inhomogeneous broadening in terms of just the first two negative-order (or, with some further approximations, positive-order) moments of any NP-size distribution function. In practice, with a simple experimental size histogram at hand, one should be immediately able to tell whether the spectra are affected by inhomogeneous broadening. We begin by considering the dipole resonance in a single metallic NP, neglecting homogeneous broadening for the moment. Such a resonance can then be described by a spectral fuction $F(\omega, R) \simeq \delta(\omega -\omega_{\mathrm{LRA}} - \eta/R)$, where $\eta (\propto \beta$ in our case) gives the strength of the leading-order $1/R$ correction associated with nonlocal response.\cite{Christensen:2014} In an ensemble of non-interacting particles characterized by a size distribution $P(R)$, the ensemble-averaged spectral function will be $\langle F(\omega) \rangle = \int dR\, F(\omega, R) P(R)$. Our aim is to express the ensemble-averaged optical properties, such as the resonance frequency $\langle \omega \rangle$, with the aid of the $n$th-order statistical moments of the particle ensemble, i.e. $\langle R^{n} \rangle = \int_{0}^{\infty} dR\, R^{n}\, P(R)$. The homogeneous delta-function line shape allows to express the $n$th-order spectral moment $\langle \omega^{n} \rangle = \int d\omega\, \omega^{n} \langle F(\omega) \rangle$ directly in terms of moments of the particle-size distribution,
\begin{equation}\label{Eq:OmegaMoments}
\langle \omega^{n} \rangle = \int dR\,  \left(\omega_{\mathrm{LRA}} + \eta/R \right)^{n} P(R) = 
\langle \left(\omega_{\mathrm{LRA}} + \eta/R \right)^{n} \rangle.
\end{equation}
It is then straightforward to derive expressions for $\langle\omega \rangle$ and the inhomogeneous broadening width, $\Delta \omega_{\mathrm{inhom}} = \sqrt{\langle \omega^{2} \rangle - \langle \omega \rangle^{2}}$, through the statistical moments of the particle-size distribution. As a key result, which allows to estimate the inhomogeneous broadening only in terms of the first two statistical moments of $P(R)$ and the nonlocal blueshift $\delta \omega_{\mathrm{LRA \rightarrow NL}} = \langle \omega \rangle - \omega_{\mathrm{LRA}} = \eta \langle R^{-1} \rangle$ ($\simeq \eta/ \langle R \rangle$ in a more crude approximation), it is shown that (see Supporting Information)
\begin{equation}\label{Eq:Broadening}
\frac{\Delta \omega_{\mathrm{inhom}}} {\delta\omega_{\mathrm{LRA \rightarrow NL}}} = 
\sqrt{\frac{\langle R^{-2} \rangle - \langle R^{-1} \rangle^{2}} {\langle R^{-1} \rangle^{2}}} 
\simeq \sqrt{\frac{\langle R^{2} \rangle - \langle R \rangle^{2}}{\langle R \rangle^{2}}}~.
\end{equation}
The first equality relates to the first and second negative-order moments of $P(R)$, which are quite unusual ways of characterizing a particle-size distribution -- in most other contexts the positive-order moments (such as the mean value and variance) are the ones of interest. In the Supporting Information we demonstrate the link between negative- and positive-order moments to obtain the second approximate identity in eq~(\ref{Eq:Broadening}), which links directly to the relative particle-size fluctuation $\Delta R/\langle R \rangle$. This result holds for any description beyond classical electrodynamics that gives a $1/R$ leading-order blueshift of the LRA resonance frequency. Most imporantly, it does not change if we replace $\eta$ with $-\eta$ to describe a corresponding $1/R$ redshift, so that our findings can be easily generalized to include other nonclassical effects, as anticipated above.

\begin{figure}[ht]
\includegraphics[width=1.0\columnwidth]{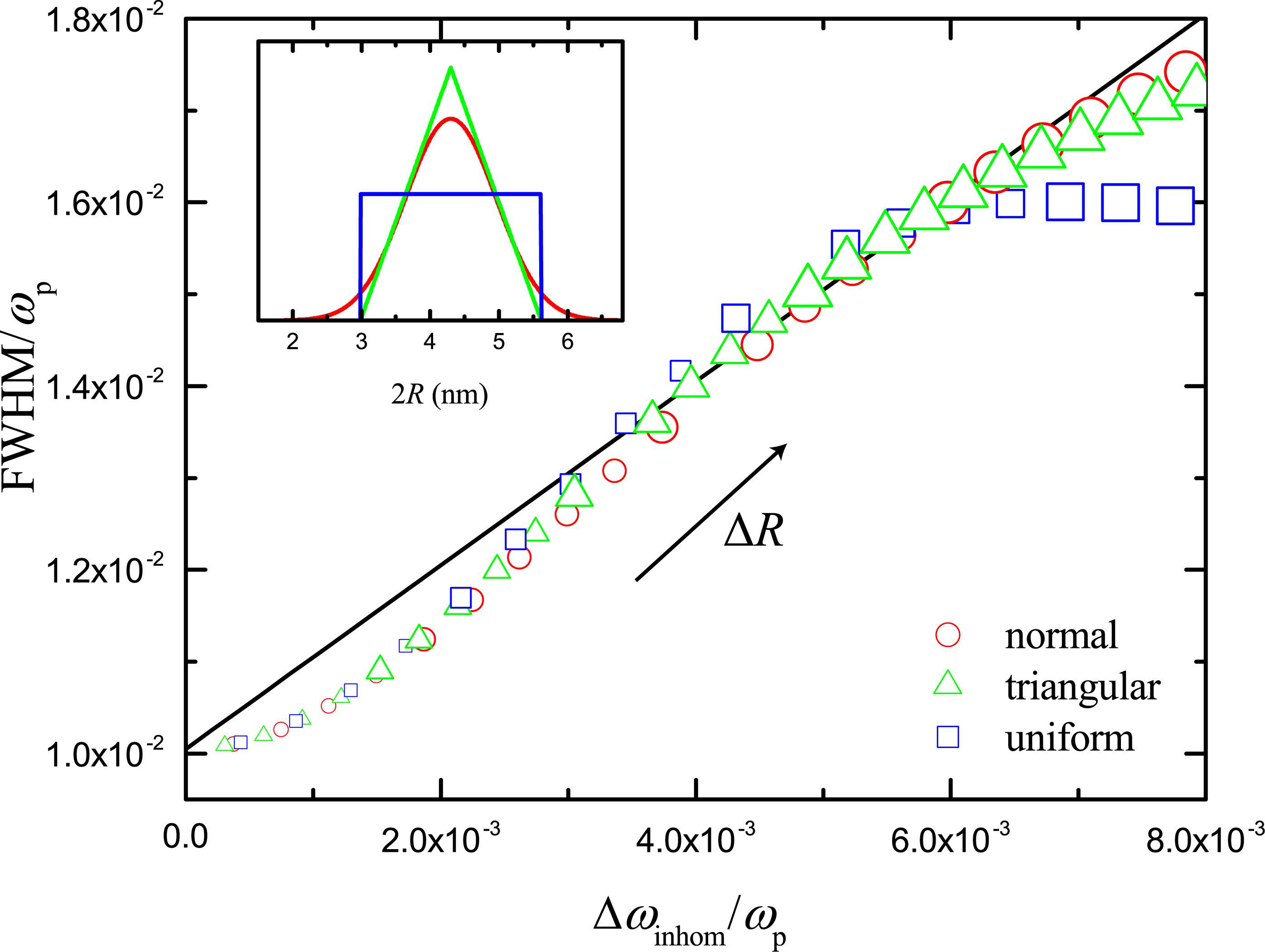}
\caption{Parametric plot (open symbols) of $\Delta \omega_{\mathrm{inhom}}$ calculated from eq~\ref{Eq:Broadening} versus FWHM obtained from simulations for a Drude-like NP within HDM ($\beta = \sqrt{3/5} v_\mathrm{F}$, $\gamma = 0.01 \omega_\mathrm{p}$ in eq~\ref{Eq:Drude}), for the size distributions shown in the inset. The average NP diameter is fixed at $2 \langle R\rangle/\lambda_{\mathrm{p}} = 0.0312$ (corresponding to 4.3 nm when $ \hbar\omega_{\mathrm{p}} = 9$ eV). Three different size distributions are plotted: uniform (blue line), triangular (green line) and (truncated) normal (red line). For the uniform (blue squares) and triangular (green triangles) cases, the distribution width increases from $0.13 \cdot 10^{-2}$ to $2.80 \cdot 10^{-2}$ (0.18 nm to 3.86 nm), while the standard deviation of the normal distribution (red circles) increases from $0.32 \cdot 10^{-3}$ to $7.00 \cdot 10^{-3}$  (0.044 nm to 0.965 nm). Increasing point size schematically depicts increasing distribution width. The black line denotes $\mathrm{FWHM} = \Delta \omega_{\mathrm{inhom}} + \mathrm{ FWHM}_{0}$.}
\label{Fig:4}
\end{figure}

To test the validity of eq~\ref{Eq:Broadening}, it is used to evaluate $\Delta \omega_{\mathrm{inhom}}$ for certain distribution shapes and widths, assuming for simplicity $\eta = \beta$. The result is then compared to the full-width-half-maximum (FWHM) of the (averaged) plasmon peak calculated in each case by simulations performed for an ideal free-electron metal within HDM, with $\beta = \sqrt{3/5} v_\mathrm{F}$ and $\gamma = 0.01 \omega_\mathrm{p}$. As long as eq~\ref{Eq:Broadening} holds, for different widths of the distribution, $\Delta \omega_{\mathrm{inhom}}$ is expected to follow a linear relation with FWHM. In Figure~\ref{Fig:4} this is done for the three distributions shown in the inset: uniform, triangular and (truncated) normal. These examples are rather extreme situations, but in all cases an almost linear relation between $\Delta \omega_{\mathrm{inhom}}$ and FWHM, following the line $\mathrm{FWHM} = \Delta \omega_{\mathrm{inhom}} + \mathrm{ FWHM}_{0}$ (black line in Figure~\ref{Fig:4}), where $\mathrm{FWHM}_{0}$ is the full-width-half-maximum of the single $\langle R \rangle$-NP, is indeed observed. For most distribution widths, all three distributions give results that lie close to this line, indicating that the simple formula of eq~\ref{Eq:Broadening} not only gives a good estimate of inhomogeneous broadening, regardless of the shape of the distribution, but can also be used to estimate the FWHM.

In summary, the effect of inhomogeneous broadening of plasmon resonances due to nonlocal response in ensembles of small plasmonic NPs was explored through detailed simulations and analytical modeling. While inhomogeneous broadening is negligible in the LRA, it can be an important issue for Drude-like metals, especially within the standard HDM approach which neglects size-dependent damping in individual NPs. Crucially, however, ensemble averaging is shown to produce almost negligible deviations in most situations of practical interest, as illustrated for realistic size distributions of noble-metal NPs, and within the more accurate GNOR model. Nanoscale experiments involving large numbers of NPs can thus be designed and analyzed in terms of the response of the mean-size NP in the ensemble, while far-field spectra of large NP collections are still expected to display the fingerprints of nonlocality, as in single-particle spectroscopies. We derived a simple equation to directly identify whether inhomogeneous broadening becomes important, simply through knowledge of the size distribution function in an ensemble. Our work provides therefore a valuable, general tool for the analysis of far-field optical spectra in modern experiments on plasmonics.

\begin{acknowledgement}
Stimulating discussions with Wei Yan are gratefully acknowledged. C.~T. was supported by funding from the People Programme (Marie Curie Actions) of the European Union's Seventh Framework Programme (FP7/2007-2013) under REA grant agreement number 609405 (COFUNDPostdocDTU).
We gratefully acknowledge support from the Villum Foundation via the VKR Centre of Excellence NATEC-II and from the Danish Council for Independent Research (FNU 1323-00087). 
The Center for Nanostructured Graphene is sponsored by the Danish National Research Foundation, Project DNRF103.
\end{acknowledgement}

\begin{suppinfo}

Derivation of eq~\ref{Eq:Broadening}, examples of its application, and short description of nonlocal Mie theory.

\end{suppinfo}

\newpage

\section{Supporting Information}

\section{Derivation of eq~\ref{Eq:Broadening}}
In the main text we have defined the inhomogeneous broadening width as
\begin{equation}
\Delta \omega_{\mathrm{inhom}} = \sqrt{\langle \omega^{2} \rangle - \langle \omega \rangle^{2}}~.
\end{equation}
From eq~(\ref{Eq:OmegaMoments}) the first- and second-order moments of $\omega$ are
\begin{equation}
\langle \omega \rangle = \langle \left(\omega_{\mathrm{LRA}} + \eta/R \right) \rangle
\end{equation}
and
\begin{equation}
\langle \omega^{2} \rangle = 
\langle \left(\omega_{\mathrm{LRA}} + \eta/R \right)^{2} \rangle,
\end{equation}
respectively. Then
\begin{eqnarray}
\Delta \omega_{\mathrm{inhom}} &=& \sqrt{\left< \left(\omega_{\mathrm{LRA}} + \frac{\eta}{R} \right)^{2} \right> -
\left< \left(\omega_{\mathrm{LRA}} + \frac{\eta}{R} \right) \right>^{2}} \nonumber\\
&=& \sqrt{ \left< \omega_{\mathrm{LRA}}^{2} + \frac{2 \omega_{\mathrm{LRA}} \eta}{R} + \frac{\eta^{2}}{R^{2}} \right> -
\left(\omega_{\mathrm{LRA}}^{2} + 2 \omega_{\mathrm{LRA}} \eta \left< \frac{1}{R} \right> + \eta^{2}\left< \frac{1}{R} \right>^{2} \right)} \nonumber \\
&=& \eta \sqrt{\left<\frac{1}{ R^2 }\right>- \left<\frac{1}{ R }\right>^2} ~.
\end{eqnarray}
In the above we have taken into account that (naturally) $\langle \omega_{\mathrm{LRA}} \rangle = \omega_{\mathrm{LRA}}$ and $\langle \eta \rangle = \eta$. Using $\delta \omega_{\mathrm{LRA \rightarrow NL}} = \eta \langle R^{-1} \rangle$, it is then straightforward to arrive to eq~(\ref{Eq:Broadening}).

\section{Statistical moments: Relating negative to positive moments}
\begin{figure}
\includegraphics[width=\linewidth]{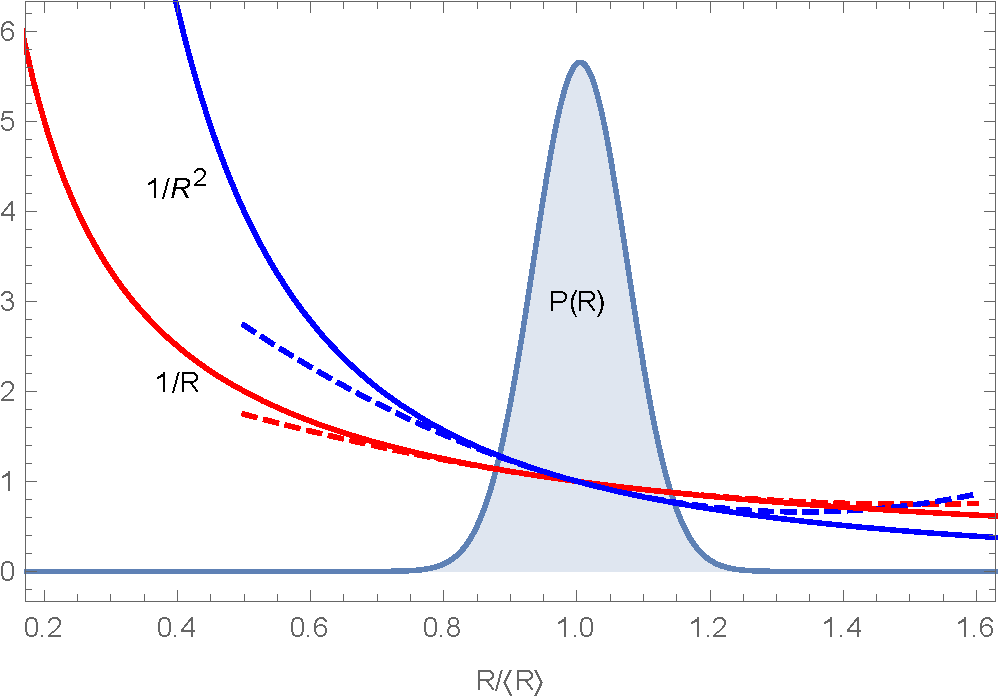}
\caption{Taylor series approximation of negative-order moments for a narrow distribution function. The dashed lines illustrate Taylor series approximations to the first and second negative moments, see eqs~\eqref{eq:x-1} and \eqref{eq:x-2}.}\label{fig:taylor}
\end{figure}

For a narrow distribution function $P(R)$, without significant small- and large-particle tails, the negative-order moments appearing in the first equality of eq.~(\ref{Eq:Broadening}) can be expressed in terms of the more common positive-order moments, to give the approximate result on the right-hand side of eq.~(\ref{Eq:Broadening}). This challenge is illustrated in Figure~\ref{fig:taylor}.

For the first negative-order moment, $R^{-1}$ can be expressed as a Taylor series expanded around the average of the distribution, $R_{0} = \langle R \rangle$:

\begin{equation}
\frac{1}{R} = \sum_{n = 0}^{+ \infty} \frac{1}{R_{0}^{n+1}} \left(R_{0}-R \right)^{n} =
\frac{1}{R_{0}} - \frac{1}{R_{0}^{2}} \left(R-R_{0} \right) + \frac{2}{2 R_{0}^{3}} \left(R-R_{0} \right)^{2} \dots ~,
\end{equation}	
\noindent
and then the moment $\langle R^{-1}\rangle$ can be calculated with:
\begin{equation}
\langle R^{-1} \rangle = \int_{-\infty}^{+\infty} R^{-1} P(R) dR = 
\int_{-\infty}^{+\infty} \sum_{n=0}^{+\infty} \frac{1}{R_0^{n+1}} \left(R_{0} - R\right)^{n} P(R) dR ~.
\end{equation}
In this expression, $R^{-1}$ and the Taylor expansion go to infinity when $R = 0$, which may cause the integral to diverge. It must therefore be required that $P(R) = 0$ for $R \leq 0$, which occurs of course for any realistic function $P(R)$. Furthermore, the summation is performed over infinite terms, and there is no immediate reason to truncate it. In fact, if $R > 2 R_{0}$ each subsequent term in the sum, $n + 1$, will be larger than the previous, $n$, and of opposite sign. To be able to truncate this series, we must ensure that each $n +1$ term is smaller than the previous one, and this is ensured by requiring that $P(R) = 0$ for $R \geq 2 R_{0}$.

We can now derive an approximate result for $\langle R^{-1} \rangle$. By including the first three terms of the series we get
\begin{equation}\label{eq:x-1}
\frac{1}{R} \simeq \frac{1}{R_{0}} - \frac{R - R_{0}}{R_{0}^{2}} + \frac{ \left(R - R_{0} \right)^{2}}{R_{0}^{3}}~,
\end{equation}
which implies that
\begin{equation}
\left< \frac{1}{R} \right> \simeq \frac{1}{R_{0}} + \frac{\langle \left(R - R_{0} \right)^{2} \rangle}{R_{0}^{3}}~,
\end{equation}
and consequently (neglecting high-order terms)
\begin{equation}
\left< \frac{1}{R} \right>^{2} \simeq \frac{1}{R_{0}^{2}} + \frac{2 \langle \left(R - R_{0} \right)^{2} \rangle}{R_{0}^{4}}~.
\end{equation}
Likewise, for the second negative moment we Taylor expand $1/R^{2}$ around $R_{0}$, to get
\begin{equation}\label{eq:x-2}
\frac{1}{R^{2}} \simeq \frac{1}{R_{0}^{2}} - \frac{2 \left(R - R_{0} \right)}{R_{0}^{3}} +
\frac{3 \left(R - R_{0} \right)^{2}}{R_{0}^{4}}~,
\end{equation}
which in turn implies that
\begin{equation}
\left< \frac{1}{R^{2}} \right> \simeq \frac{1}{R_{0}^{2}} +
\frac{3 \langle \left(R - R_{0} \right)^{2} \rangle}{R_{0}^{4}} ~.
\end{equation}
Then, for the size fluctuations we have
\begin{equation}
\left< \frac{1}{R^{2}} \right> - \left< \frac{1}{R} \right>^2 \simeq 
\frac{\langle \left(R - R_{0} \right)^{2} \rangle}{R_{0}^{4}} ~,
\end{equation}
and thus
\begin{equation}\label{eq:moments-approximation}
\sqrt{\left< \frac{1}{R^{2}} \right> - \left< \frac{1}{R} \right>^{2}} \simeq 
\frac{1}{R_{0}} \frac{\sqrt{\langle \left(R - R_{0} \right)^{2} \rangle}}{R_{0}} =
\frac{1}{\langle R \rangle}\frac{\sqrt{\langle R^{2} \rangle - \langle R \rangle^{2}}}{\langle R \rangle} ~.
\end{equation}

\section{Uniform distribution}
As a particular example that can be treated analytically, we consider a uniform distribution function
\begin{equation}
P(R) = \frac{1}{\delta R} \theta(R - R_{0} + \delta R/2) \theta(-R + R_{0} + \delta R/2)~,
\end{equation}
where $\theta(x)$ is the Heaviside function. By construction, $P(R)$ is normalized and with a mean value of $\langle R \rangle = R_{0}$, while $\langle (R - R_{0})^2 \rangle = \frac{1}{12} (\delta R)^{2}$. The requirement that all radii in the distribution are positive gives a bound on its parameters, namely that $R_{0} \ge \delta R/2$. For the first negative-order moment we get
\begin{equation}
\langle R^{-1} \rangle = \frac{1}{\delta R} \int_{R_{0}-\delta R/2}^{R_{0}+\delta R/2} dR\, R^{-1} = 
\langle R \rangle^{-1} g_{1} (\tfrac{\delta R}{\langle R \rangle})~,
\end{equation}
with
\begin{equation}
g_{1}(x) = x^{-1} \ln \left(\frac{2+x}{2-x}\right) = 1 + \frac{1}{12} x^{2} + {\mathcal O}(x^{4})~.
\end{equation}
Similarly, for the second-order negative moment we get
\begin{equation}\label{eq:minus_second_moment_uniform}
\langle R^{-2} \rangle = \frac{1}{\delta R} \int_{R_{0} - \delta R/2}^{R_{0} + \delta R/2} dR\, R^{-2} = 
\langle R \rangle^{-2} g_{2} (\tfrac{\delta R}{\langle R \rangle})~,
\end{equation}
with
\begin{equation}
g_{2}(x) = \frac{4}{4 - x^{2}} = 1 + \frac{1}{4} x^{2} + {\mathcal O}(x^{4}).
\end{equation}
In this way we can directly calculate
\begin{equation}
\sqrt{\left< \frac{1}{R^{2}} \right> - \left< \frac{1}{R} \right>^{2}} \simeq 
\frac{1}{\sqrt{12}} \frac{\delta R}{R_{0}^{2}}~.
\end{equation}
Returning to eq~\ref{eq:moments-approximation} we indeed find the same result. In a similar way, for the case of the triangular and normal distribution that concern us in the main text, the result is
\begin{equation}
\text{Triangular:} \qquad \sqrt{\left< \frac{1}{R^{2}} \right> - \left< \frac{1}{R} \right>^{2}} \simeq 
\frac{1}{\sqrt{24}} \frac{\delta R}{R_{0}^{2}}
\end{equation}
\begin{equation}
\text{Normal:} \qquad \sqrt{\left< \frac{1}{R^{2}} \right> - \left< \frac{1}{R} \right>^{2}} \simeq 
\frac{\sigma}{R_{0}^{2}}~,
\end{equation}
where $\sigma$ is the standard deviation of the normal distribution. Note that the normal distribution is truncated, limited in the region $R = 0 - 2 R_{0}$.

\section{Nonlocal Mie theory}

Here we summarize the fully-retarded Mie theory for a spherical plasmonic particle treated within HDM. The multipolar response of a sphere including nonlocal effects was determined by Ruppin\cite{Ruppin:1973,Ruppin:1975} by extending Mie theory\cite{Bohren:1983} to take into account excitation of longitudinal waves. In the framework of Mie theory, the extinction cross section of a sphere of radius $R$ embedded in a homogeneous host medium is given by\cite{Bohren:1983}
\begin{equation}\label{eq:extinction_ret}
\sigma_{\mathrm{ext}} = -\frac{2\pi}{q_{\mathrm{h}}^{2}} \sum_{\ell = 1}^{+\infty} \left(2\ell + 1\right) \mathrm{Re} \left(t_{\ell}^{TE} + t_{\ell}^{TM} \right)~, 
\end{equation}
where $\ell$ denotes the angular momentum and $q_{\mathrm{h}}$ is the wavenumber in the host medium, which is described by a dielectric function $\varepsilon_{\mathrm{h}}$. Assuming that the magnetic permeabilities, both in the sphere and in the host medium are equal to 1, the nonlocal Mie scattering coefficients are\cite{Ruppin:1973,Ruppin:1975,David:2011,Christensen:2014}
\begin{subequations}\label{eq:Mie_coefs}
\begin{align}
t_{\ell}^{TE} & = 
\frac{-j_{\ell}(x_{\mathrm{t}}) [x_{\mathrm{h}} j_{\ell} (x_{\mathrm{h}})]' + j_{\ell} (x_{\mathrm{h}}) [x_{\mathrm{t}} j_{\ell}(x_{\mathrm{t}})]'}
{j_{\ell} (x_{\mathrm{t}}) [x_{\mathrm{h}} h_{\ell}^{+} (x_{\mathrm{h}})]'- h_{\ell}^{+}(x_{\mathrm{h}}) [x_{\mathrm{t}} j_{\ell}(x_{\mathrm{t}})]'}, \label{eq:Miecoefs_TE}\\
t_{\ell}^{TM} & = 
\frac{-\varepsilon_{\mathrm{t}} j_{\ell} (x_{\mathrm{t}}) [x_{\mathrm{h}} j_{\ell} (x_{\mathrm{h}})]' + \varepsilon_{\mathrm{h}} j_{\ell} (x_{\mathrm{h}})\left\{ [x_{\mathrm{t}} j_{\ell} (x_{\mathrm{t}})]' + \Delta_{\ell} \right\} }
{\varepsilon_{\mathrm{t}} j_{\ell} (x_{\mathrm{t}}) [x_{\mathrm{h}} h_{\ell}^{+} (x_{\mathrm{h}})]' - \varepsilon_{\mathrm{h}} h_{\ell}^{+} (x_{\mathrm{h}})\left\{ [x_{\mathrm{t}} j_{\ell} (x_{\mathrm{t}})]' + \Delta_{\ell} \right\} }~,\label{eq:Miecoefs_TM}
\end{align}
where $j_{\ell} (x)$ and $h_{\ell}^{+} (x)$ are the spherical Bessel function and Hankel function of the first type, respectively, while
$x_{\mathrm{h}} = q_{\mathrm{h}} R$ and $x_{\mathrm{t}} = q_{\mathrm{t}} R$. Here $q_{\mathrm{t}}$ is the (transverse) wavenumber inside a sphere described by a transverse dielectric function $\varepsilon_{\mathrm{t}}$. The nonlocal correction $\Delta_{\ell}$ to the Mie coefficients is given as
\begin{equation}\label{eq:CNL}
\Delta_{\ell} = \ell \left(\ell + 1\right) j_{\ell} (x_{\mathrm{t}}) \frac{\varepsilon_{\mathrm{t}} - \varepsilon_{\infty}}{\varepsilon_{\infty}} \frac{j_{\ell} (x_{\mathrm{l}})}{x_{\mathrm{l}}j_{\ell}'(x_{\mathrm{l}})} ~, 
\end{equation}
\end{subequations}
where $x_\mathrm{l} = q_{\mathrm{l}} R$ and $q_{\mathrm{l}}$ is the longitudinal wavenumber in the sphere, associated with the longitudinal dielectric function $\varepsilon_{\mathrm{l}}$, which is frequency- and wavevector-dependent. The dispersion of longitudinal waves is given by $\varepsilon_{\mathrm{l}} (\omega, \mathbf{q}) = 0$. In the limiting case where $\Delta_{\ell} = 0$ we retrieve the local result of standard Mie theory. All our numerical results in the main text have been obtained from numerical evaluations of eq~\eqref{eq:extinction_ret}.

\bibliography{Nonlocal_inhom_broadening}

\end{document}